\documentstyle[epsfig,prl,aps]{revtex}
\newcommand{\doublespace}{
  \renewcommand{\baselinestretch}{1.75}
  \large\normalsize}
\begin{document}
\input epsf.sty
\def\Im{\mbox{Im}}
\doublespace
\large

\vspace{.2in}

\title{Physical Mechanism of the $d\rightarrow d+is$ Transition}
\author{\v Simon Kos}
\address{University of Illinois, Department of Physics\\ 1110 W. Green St.\\
Urbana, IL 61801 USA
\\E-mail: s-kos@uiuc.edu
}

\maketitle

\begin{abstract}

We discuss the basic physical mechanism of the $d\rightarrow d+is$ transition,
which is the currently accepted explanation for the results of tunneling
experiments into $ab$ planes \cite{covington97}. Using the first-order
perturbation theory, we show that the zero-bias states drive the transition.
We present various order-of-magnitude estimates and consistency checks that
support this picture.

\end{abstract}

\vspace{.2in}

It has now been firmly established \cite{dale95} 
that the order parameter in cuprate 
superconductors has a $d$-wave symmetry. An inhomogeneity 
may therefore scatter a quasiparticle between directions that
experience opposite signs of the order parameter. This effect is strongest
for specular reflection off the 110 surface. 
Here the sign of the order 
parameter along every quasiclassical trajectory changes sign upon reflection
(see Fig. 1). As a consequence of this and 
the Atiyah-Patodi-Singer index theorem \cite{atiyah}, the
Andreev equation along each trajectory has a normalizable eigenstate of zero
energy \cite{hu}.
These all contribute to the local density of states, and are the source
of the zero-bias anomaly observed in
tunneling experiments \cite{geerk,lesueur,covington96}. 

The tunneling data \cite{covington97} show that the zero-bias peak splits as 
the temperature is lowered below 1 or 2 K.
This splitting is probably due to an additional symmetry-breaking transition,
most likely $d \rightarrow d+is$ transition where a subdominant $s$-wave
order parameter appears close to the surface with a $\pi /2$ phase shift
compared to the dominant $d$-wave. Such a transition had been anticipated
theoretically \cite{matsumoto} prior to the experiment.
Subsequent calculations of the surface phase diagram \cite{fogelstrom,rainer}
included the effects of the surface roughness, and gave results that are
in good agreement with the experimental data.

These calculations are done within the Eilenberger approach to
superconductivity \cite{eilenberger}, which uses the quasiclassical 
approximation of the local Green function.
This (so-called Eilenberger) function then satisfies a transport-like
equation. Unfortunately, this approach is very formal, so it hides,
rather than explains, the physics behind the calculations, especially when
the equations are solved numerically in the imaginary-time domain
\cite{matsumoto,buchholtz}. As new
experimental results \cite{dale}
call the $d+is$ interpretation into question, it is even
more important to have a direct physical understanding of the transition.

We believe the Andreev quasiclassical
picture \cite{andreev}, which deals with single-particle states rather than
local Green functions, contributes to such understanding by showing 
how the energy
costs and benefits of the $d\rightarrow d+is$ transition are distributed among
the degrees of freedom in the system. As we mentioned above, due to the
presence of the 110 surface, there are many zero-bias states (ZBSs) in the
spectrum. The main purpose of this paper is to show that these states drive
the transition.
\vskip 5 mm
\begin{figure}
\epsfig{file=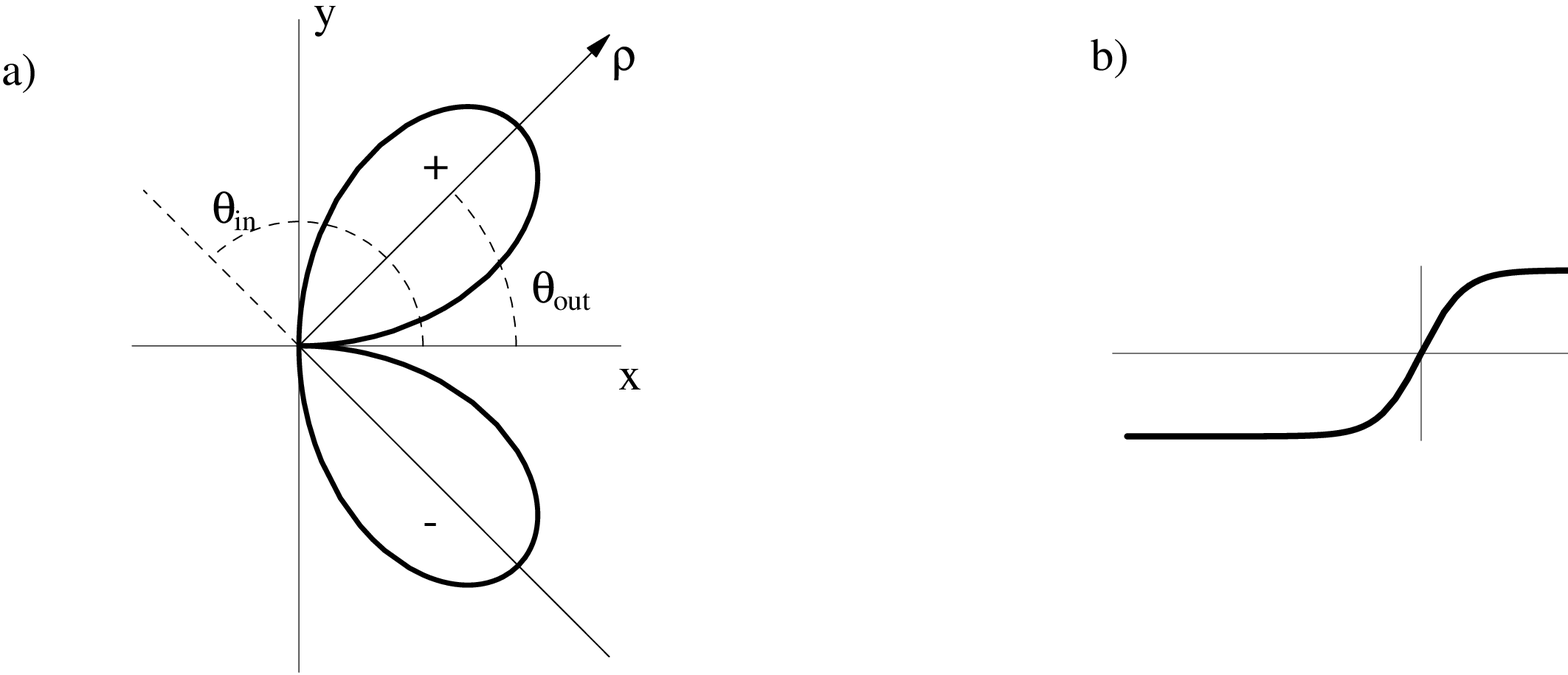,width=12.0cm}
\caption{a) A schematic picture of the normal metal--superconductor junction in
the 110 direction with a typical quasiclassical trajectory.
\newline
b) A schematic graph of the pairing potential along the trajectory in a).}
\end{figure}

Our approach was inspired by the study of a weak-coupling BCS superconductor
modeled by a Fermi sea with an attractive local four-fermion interaction
$$
V\psi _{\uparrow}^{\dag}({\bf r}) \psi _{\downarrow}^{\dag}({\bf r})
\psi _{\downarrow}({\bf r}) \psi _{\uparrow}({\bf r});
$$
$V<0$ is the coupling constant. One decouples this interaction by the 
Hubbard-Stratonovich (HS) transformation, which introduces the pairing field
$\Delta $. The instability to superconductivity can then be detected by going
to $T=0$ and looking at the energy change due to the opening of the gap 
$|\Delta |$. The occupied single-particle states lower their energy as
$|\Delta |^2 \ln |\Delta |$ while the system raises its energy by
$|\Delta |^2 /(-V)$, which is the extra term in the hamiltonian introduced by
the HS transformation. Since the non-analytic decrease wins over the analytic
increase as $|\Delta |\rightarrow 0$, an arbitrarily weak attractive $V$ will,
at low enough temperature, make the system superconducting. The decrease
of the single-particle energies is  non-analytic due to contribution from the
states initially close to the Fermi surface. We conclude, therefore, that 
these low-energy states drive the BCS transition.

A similar argument works for the $d\rightarrow d+is$ transition. If there
is an attractive interaction in the $s$-wave channel with strength $V_s<0$,
we may use the HS transformation to introduce the $s$-wave component of the
pairing field $\Delta _s$ on top of the dominant $\Delta _d$, leading to
an extra positive term in the hamiltonian $|\Delta _s|^2 /(-V_s)$, just as
in the BCS case. When the ``$is$'' component appears close to the surface, the
shape of the order parameter along the trajectory in Fig. 1 changes, as shown
in Fig. 2. To study the instability to the transition, we again look at
$T=0$ and $|\Delta _s|\rightarrow 0$. The energy of the
ZBSs \footnote{We shall abuse the terminology and call these states ZBSs
even after their energy has been shifted away from zero.} to the lowest,
that is first,
order in $\Delta _s$ then changes to \cite{long}
\begin{equation}
\label{EZBS}
E_{\theta }[\Delta _s] = \pm \int\limits_{-\infty}^{+\infty} d\rho
2|f(\theta ,\rho )|^2 \Im \Delta _s,
\end{equation}
where the upper (lower) sign corresponds to the up- (down-)moving trajectory.
Here
$$
\pmatrix{f(\theta ,\rho ) \cr g(\theta ,\rho )}
$$
is a solution of the Andreev equation along the trajectory in the direction
$\theta $ at the point $\rho $. For the ZBSs, 
$g(\theta ,\rho ) = \mp i f(\theta ,\rho )$. At $T=0$, only the down-moving
ZBSs will be occupied, and their energy will decrease linearly. It turns out
\cite{long} that the energy changes of the remaining states (non-ZBSs) cancel
each other out. Thus, upon the appearance of small $|\Delta _s|$, the energy
of the occupied (zero-bias) states is lowered linearly, which, for small 
enough $|\Delta _s|$, wins over the quadratic increase of the HS term for
arbitrarily weak $s$-wave attraction $V_s$. It follows that the transition
$d\rightarrow d+is$ is driven by the ZBSs.
\vskip 5mm
\begin{figure}
\epsfig{file=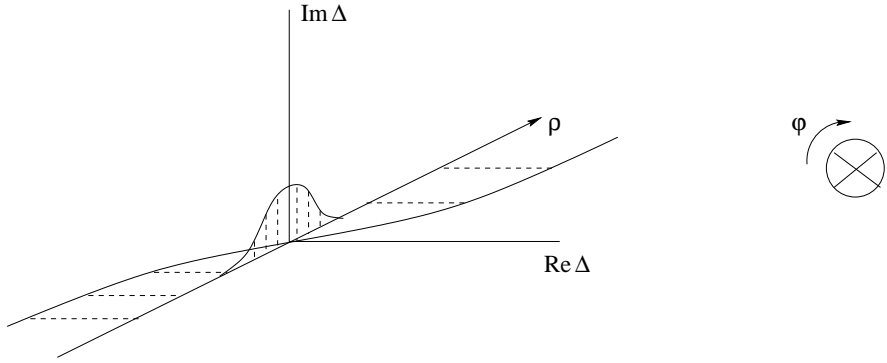,width=12.0cm}
\caption{The pairing potential along the trajectory in Fig. 1a). The 
corresponding twist of the phase of the order parameter is clockwise.}
\end{figure}

For YBCO, the experiments give $|\Delta _d| \sim 30$ meV (amplitude) and
$|\Delta _s|\sim 1$ meV, so we assume the first-order perturbation formula
(\ref{EZBS}) holds up to the experimental value of $|\Delta _s|$. We can then
calculate $|\Delta _s|$ by minimizing the energy of the system. When we sum
over all occupied ZBSs, we find that the energy per CuO plane per unit
length of the surface is
\begin{equation}
\label{totE}
E[s(x)] = \int\limits_0^{\infty } dx{s^2(x) \over (-V_s)} +
\int\limits_{-\pi /2} ^0 {k_F\over 2\pi } d\theta \cos \theta
E_{\theta }[s(x)],
\end{equation}
where $s(x) = \Im \Delta _s (x)$. Since $s$ extends into the bulk only as
far as the $d$-wave coherence length $\xi \equiv \hbar v_F/|\Delta _d|$, we
can estimate
\begin{equation}
E[s] \sim {s^2 \over (-V_s)} \xi - k_F s,
\end{equation}
which gives
\begin{equation}
\label{sestim}
s\sim {(-V_s) k_F \over \xi}.
\end{equation}
The experimental values $s\sim 1$ meV, $k_F \sim 1 $ \AA $^{-1}$, and
$\xi \sim 10 $ \AA \  give us {$|V_s| \sim 10 $ meV \AA $^2$}.
(Since the CuO plane is two-dimensional, $V_s$ has dimension
$EL^2$ rather than $EL^3$.)

The full variational calculation is presented in \cite{long}. We only remark
here that the solution of the variational equation obtained from (\ref{totE})
agrees with the contribution to $\Delta $ from the occupied ZBSs in the
gap equation
\begin{equation}
\label{gapeqn}
\Delta _{ZBS}(x)= i (-V_s)\int\limits_{-\pi /2}^0 {k_F\over 2\pi} d\theta
2|f(\theta ,x/\cos \theta)|^2,
\end{equation}
that is, $\Delta _{ZBS}(x)= is(x)$, which shows the internal consistency of the 
picture. Unlike the BCS gap equation, (\ref{gapeqn}) is an explicit formula
for $\Delta _{ZBS}(x)$; there is no $\Delta _{ZBS}$ on the right-hand side.
The physical reason for this is that within the first-order perturbation
theory, it is only the sign of $s(x)$ (and not its magnitude) that determines
which ZBSs are occupied.

Our argument has shown only that $d+is$ is the favorable state at $T=0$. To
see what happens at finite temperatures, we need to minimize the free energy
$F[s(x)]$, which is obtained from (\ref{totE}) when we replace 
$E_{\theta }[s(x)]$ by $(-T) \ln (1 + \exp (-E_{\theta }[s(x)] /T))$,
making the lowest order quadratic in $s(x)$. The order-of-magnitude estimate
gives
\begin{equation}
F[s] \sim \left( {\xi \over (-V_s)} - {k_F \over T} \right)s^2
+ O(s^4).
\end{equation}
Hence, the system is unstable to the transition to the $d+is$ state even at
finite temperatures. The transition temperature is
\begin{equation}
T_s \sim {(-V_s) k_F \over \xi },
\end{equation}
which is of the same order of magnitude as $|\Delta _s|_{T=0}$ 
(see (\ref{sestim})).

Now we shall go back to $T=0$ and observe that the presence of the $is$ 
component induces a twist in the phase $\varphi $ of the order parameter
from $-\pi $ to 0 for an up-moving trajectory and from 0 to $\pi $ for a
down-moving one, see Fig. 2. This twist 
implies a current flowing down; its density
at the point $\rho $ of a quasiclassical trajectory labeled by $\theta $ is
equal to
\begin{equation}
j(\theta ,\rho )={e\over 2m} n^{(1d)} \partial _{\rho } \varphi (\theta ,\rho ),
\end{equation}
where $n^{(1d)} \equiv k_F/\pi $ is the density of the one-dimensional Fermi
sea ($k_F$ is the Fermi wave vector). Current flowing down is also expected from
our previous argument that showed that only the down-moving ZBSs are occupied.
Each of them contributes
\begin{equation}
j(\theta ,\rho )= ev_F (|f(\theta ,\rho )|^2 + |g(\theta ,\rho )|^2)
\end{equation}
to the current density at a given point; it turns out again \cite{long} that
the contributions from the non-ZBSs cancel. In both approaches, we get the
total current by summing up contributions from all the quasiclassical
trajectories. The result is
\begin{equation}
I={ev_Fk_F\over 4\pi}
\end{equation}
per CuO plane in both calculations. To get an order-of-magnitude estimate,
we take $v_F \sim 10^5 $ m/s and $k_F \sim 10^{10} $ m$^{-1}$, so
$I\sim 10^{-5} $ A. The agreement shows that all the current is carried by
the ZBSs. It is important that not only the magnitude, but also the direction
of the current, agrees in both calculations. The reason for this agreement is
that
the ZBSs moving in the direction of the current are shifted down in energy
and thus occupied, whereas those moving against the current are shifted up 
and unoccupied. We wish to stress that this is exactly opposite to the sign
of the Doppler shift: the states moving along the current would be
Doppler-shifted up, whereas those moving against the current would be 
Doppler-shifted down.

In conclusion, the first-order perturbation theory shows that the 
$d\rightarrow d+is$ transition is driven by the ZBSs and that the transition
temperature is finite and of the order $|\Delta _s|_{T=0}$. The ZBSs that
are being pushed down by the $is$ component carry surface current; they are
{\bf not } Doppler-shifted by this current.

I want to give special thanks to M. Stone for suggesting the problem and
discussing it with me. I have also benefited from discussions with 
H. Aubin, L. H. Greene, A. J. Leggett, D. E. Pugel, R. Ramazashvili, 
M. Turlakov,
and H. Westfahl. I would like to thank C. M. Elliott for proofreading the
manuscript. The project was supported by the grant NSF-DMR-98-17941.

\end{document}